\documentclass[final,3p,times]{elsarticle}

\usepackage{amssymb}
\usepackage{amsmath}
\usepackage{multirow}

\newcommand\mat\mathbf
\newcommand\dd{\, \mathrm{d}}
\newcommand\ii{\mathrm{i}}

\newcommand\um{~{\mu}\textrm{m}}
\newcommand\nm{\textrm{~nm}}

\makeatletter
\def\ps@pprintTitle{%
  \let\@oddhead\@empty
  \let\@evenhead\@empty
  \let\@oddfoot\@empty
  \let\@evenfoot\@oddfoot
}
\makeatother

\begin{document}

\begin{frontmatter}


\title{Vectorial spatial differentiation of optical beams with metal-dielectric multilayers enabled by spin Hall effect of light and resonant reflection zero}

\author[label1,label2]{Leonid~L.~Doskolovich}
\author[label1,label2]{Artem~I.~Kashapov}
\author[label1,label2]{Evgeni~A.~Bezus}
\author[label1,label2]{Dmitry~A.~Bykov}

\affiliation[label1]{organization={Samara National Research University},
 addressline={34 Moskovskoye shosse}, 
 city={Samara},
 postcode={443086}, 
 country={Russia}}
						
\affiliation[label2]{organization={Image Processing Systems Institute, National Research Centre ``Kurchatov Institute''},
 addressline={151 Molodogvardeyskaya st.}, 
 city={Samara},
 postcode={443001}, 
 country={Russia}}

\begin{abstract}
We theoretically describe and numerically investigate the operation of “vectorial” optical differentiation of a three-dimensional light beam, which consists in simultaneous computation of two partial derivatives of the incident beam profile with respect to two spatial coordinates in different transverse electric field components.
It is implemented upon reflection of the beam from a layered structure by simultaneously utilizing the effect of optical resonance and the spin Hall effect of light. As an example of a layered structure performing this operation, we propose a three-layer metal-dielectric-metal (MDM) structure.
We show that by choosing the parameters of the MDM structure, it is possible to achieve the so-called isotropic vectorial differentiation, for which the intensity of the reflected optical beam (squared electric field magnitude) is proportional to the squared absolute value of the gradient of the incident linearly polarized beam.
The presented numerical simulation results demonstrate high-quality vectorial differentiation and confirm the developed theoretical description.
\end{abstract}

\begin{keyword}
optical differentiation \sep layered structure \sep resonance \sep spin Hall effect of light \sep transfer function
\end{keyword}

\end{frontmatter}

\section{Introduction}
In recent years, the development of photonic structures for analog optical computing has attracted great interest.
These structures are considered as a new promising platform that will help overcoming the limitations of digital computing systems associated with insufficient computing speed and high energy consumption~\cite{1,2,3}.
Much attention in the field of analog optical computing is paid to the development of compact photonic structures for the differentiation of optical signals (with respect to time or spatial coordinates), as well as for the implementation of more complex differential operators~\cite{1,2,3,4,5,6,7,8,9,10,11,12,13,14,15,16,17,18,19,20}.

A necessary condition for optical differentiation is the presence of a zero in the reflection or transmission spectrum of the diffractive structure being used. 
Due to the fact that the appearance of zeros in the reflection (transmission) spectrum of a photonic structure is often associated with the excitation of its eigenmodes, various resonant structures are widely used for optical differentiation, e.\,g., resonant gratings (photonic crystal slabs)~\cite{3,4,5,6,7,8,9}, resonant layered structures and their integrated counterparts~\cite{10,11,12,13,14,15,16,17}, and micro-resonators~\cite{19,20,21}.
In addition to resonant structures, simple non-resonant structures have also been proposed for performing optical differentiation, including structures utilizing the Brewster effect~\cite{22,23} and the optical analogue of the spin Hall effect~\cite{24,25,26}.
Besides, it was shown that a pair of two properly oriented polarizers enables obtaining a reflection zero required for differentiation~\cite{27,28}.
It is also important to note that, along with analog computing itself, photonic structures for spatial differentiation (i.\,e., for differentiating the transverse profile of an incident optical beam with respect to one or two spatial variables) are promising for image processing systems, since they enable performing real-time optical edge detection in input images~\cite{3,14,15,22,23,24,25,26,27,28}.

One of the simplest resonant structures that can be used for optical differentiation is a three-layer metal-dielectric-metal (MDM) structure consisting of two metal layers separated by a dielectric layer~\cite{11,12,13}.
In such a structure, it is possible to obtain a reflection zero due to the excitation of a quasiguided mode localized in the dielectric layer. 
In particular, in the previous works, the present authors considered the use of MDM structures operating in the oblique incidence geometry for first-order differentiation with respect to time or one spatial coordinate~\cite{11,12}, as well as for spatiotemporal differentiation corresponding to the calculation of a weighted sum of the partial derivatives with respect to time and a spatial coordinate~\cite{13}.
It is interesting to note that the transfer function that describes the operation of the MDM structure as a spatiotemporal differentiator~\cite{13} has a zero possessing a topological charge of $\pm1$.
In this sense, the MDM structure is a topological differentiator~\cite{27}, which is robust to small deformations caused, for example, by errors in the layer thicknesses appearing at the fabrication stage.
In addition to first-order differentiation, it was shown that a properly designed “double” MDM structure (i.\,e., a structure consisting of two three-layer MDM structures separated by a dielectric layer) possesses a second-order reflection zero and enables optically calculating the Laplace operator in the oblique incidence geometry~\cite{14}.
This makes it promising for the edge detection (contour enhancement) problem~\cite{14}.
In~\cite{29}, to enhance contours, a single (three-layer) MDM structure was used, which implemented the optical computation of a weighted sum of second-order partial derivatives with respect to two spatial variables at normal incidence.
At the same time, the weights of the second derivatives in~\cite{29} were significantly different, which led to essentially “non-isotropic” edge detection.

It is important to note that in optical differentiators based both on resonant MDM structures (as well as other layered structures) and on more complex resonant structures including resonant diffraction gratings and metasurfaces, the differentiation operation is carried out in one of the electric field components of the resulting (reflected or transmitted) beam either coinciding with the “main” field component of the incident optical signal~\cite{1,3,4,5,6,7,8,9,10,11,12,13,14,15,16,17,18,19,20,21,22,23} or being “cross-polarized” with respect to the main component~\cite{24,25,26}.
To the best of our knowledge, in the existing works, the operation of simultaneous calculation of two derivatives with respect to different variables in two different components of the electric field of a reflected (or transmitted) beam has not yet been systematically studied.
Such an operation can be referred to as the “vectorial” differentiation. 
This operation is of great interest, especially in the case, in which the magnitudes of the coefficients (weights) of the two calculated partial derivatives coincide. 
In this case, the intensity of the generated light beam (the squared magnitude of the electric field) will be proportional to the squared absolute value of the gradient of the electric field of the incident linearly polarized beam. 
In the opinion of the present authors, the operation of calculating the squared gradient is of great practical importance, since it can be used for “isotropic” edge detection in an input image (incident beam), i.\,e., for equally effective enhancement of contours having different orientations. 
In this regard, the operation of vectorial differentiation with equal absolute values of the coefficients (weights) of the calculated partial derivatives can be referred to as the operation of isotropic vectorial differentiation.

In this work, we develop a systematic theoretical description of the operation of vectorial differentiation using a layered structure for a three-dimensional incident optical beam. 
It is shown that the simultaneous utilization of the optical resonance effect, which provides a reflection zero, and the optical analogue of the spin Hall effect makes it possible to perform vectorial differentiation, i.\,e., to compute two different partial derivatives in two different transverse components of the electric field of the reflected beam. 
As an example of a layered structure implementing this operation, we propose a three-layer MDM structure operating at oblique incidence. 
It is shown that by choosing the parameters of the MDM structure, it is possible to carry out isotropic vectorial differentiation, which makes this simple structure efficient in the problem of optical edge detection.

\section{Diffraction of a three-dimensional optical beam on a layered structure}\label{sec:1}
\subsection{Representation of the incident beam}

\begin{figure}[hbt]
	\centering
		\includegraphics{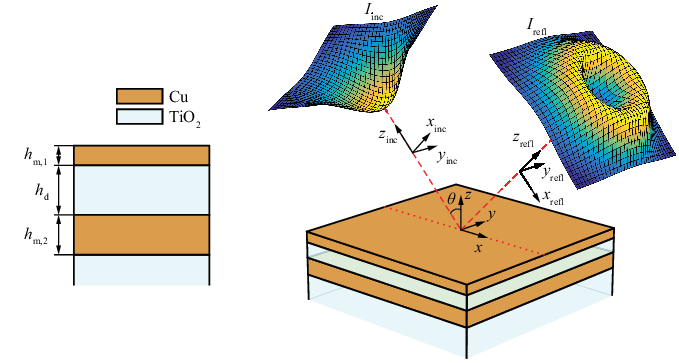}
	\caption{\label{fig:1} Geometry of the problem of diffraction of a three-dimensional optical beam on a layered structure (right) and schematic cross-section of the utilized three-layer MDM structure (left)}
\end{figure}

Following Ref.~\cite{14}, to describe the diffraction of a monochromatic light beam obliquely incident on a layered structure at an angle of incidence $\theta$ (Fig.~\ref{fig:1}), let us represent the incident beam in the basis of plane waves with transverse electric (TE) and transverse magnetic (TM) polarizations written in the coordinate system $(x_{\rm inc}, y_{\rm inc}, z_{\rm inc})$, which is rotated by an angle $\theta$ measured from the normal to the interfaces of the structure and is associated with the incident beam. 
As it is shown in Fig.~\ref{fig:1}, in this coordinate system, the incident beam propagates in the negative direction of the $z_{\rm inc}$ axis. 
In order to obtain the equations describing the field of TM- and TE-polarized plane waves in the coordinate system of the incident beam, let us first write these waves in the global coordinate system $(x, y, z)$ associated with the structure (see Fig.~\ref{fig:1}), and then perform a rotation transformation to move to the coordinate system $(x_{\rm inc}, y_{\rm inc}, z_{\rm inc})$.

In the global coordinate system ${\mat{r}} = (x,y,z)$, the equations of TM- and TE-polarized plane waves having the wave vector $\mat{k} = (k_x, k_y, k_z)$ can be obtained from Maxwell’s equations as~\cite{14, 30}
\begin{equation}\label{eq:1}
\begin{aligned}
 \begin{bmatrix}
 E_x \\ E_y \\ E_z \\ H_x \\ H_y \\ H_z
\end{bmatrix} 
&= \mat{\Psi}_{\rm TM}(\mat{k})\exp \{\ii \mat{k} \cdot \mat{r}\} 
= \frac{1}{{\sqrt{k_x^2 + k_y^2}}}
\begin{bmatrix}
 - k_x k_z \\ 
 - k_y k_z \\ 
 k_x^2 + k_y^2 \\ 
 k_0 k_y \varepsilon_{\rm sup} \\ 
 -k_0 k_x \varepsilon_{\rm sup} \\ 
 0
\end{bmatrix} \exp \left\{\ii k_x x + \ii k_y y + \ii k_z z \right\},
\\
\begin{bmatrix}
 E_x \\ E_y \\ E_z \\ H_x \\ H_y \\ H_z
\end{bmatrix} 
&=\mat{\Psi}_{\rm TE}(\mat{k})\exp \{\ii \mat{k} \cdot \mat{r}\} 
= \frac{1}{{\sqrt{k_x^2 + k_y^2}}} 
\begin{bmatrix}
 -k_0 k_y \\ 
 k_0 k_x \\ 
 0 \\ 
 -k_x k_z \\ 
 -k_y k_z \\ 
 k_x^2 + k_y^2
\end{bmatrix}\exp \left\{\ii k_x x + \ii k_y y + \ii k_z z \right\}.
\end{aligned}
\end{equation}
Note that the components of the wave vector $\mat{k} = (k_x, k_y, k_z)$ satisfy the dispersion relation $k_x^2 + k_y^2 + k_z^2 = k_0^2 \varepsilon_{\rm sup}$, where $k_0 = 2\pi / \lambda$ is the wave number, $\lambda$ is the free-space wavelength, and $\varepsilon_{\rm sup}$ is the dielectric permittivity of the medium. 
In the case of a TM-wave [first equation in Eq.~\eqref{eq:1}], the magnetic field vector is perpendicular to the plane of incidence containing the wave vector $\mat{k}$ and the $z$ axis, whereas the electric field vector lies in this plane. 
For a TE-wave, the electric field vector is perpendicular to the plane of incidence and the magnetic field vector lies in this plane. 
A wave with an arbitrary polarization can be represented by a superposition of TM- and TE-waves.

As it was noted above, for representing a beam incident on the structure at an angle of incidence $\theta$ (Fig.~\ref{fig:1}), we will use a coordinate system $\mat{r}_{\rm inc} = (x_{\rm inc}, y_{\rm inc}, z_{\rm inc})$ associated with the incident beam, which can be obtained from the global coordinate system by a rotation by an angle $\theta$ about the $y$ axis. 
In this case, the coordinates of the vectors in the two considered coordinate systems will be related by the following linear transformation:
\begin{equation}\label{eq:2}
\mat{r}_{\rm inc}^{{\rm T}} 
= \mat{R}_y \mat{r}^{\rm T} 
= \begin{bmatrix}
 \cos\theta & 0 & \sin \theta \\ 
 0&1&0 \\ 
 -\sin\theta & 0 & \cos\theta
\end{bmatrix}
\begin{bmatrix}
 x_{\rm inc} \\ 
 y_{\rm inc} \\ 
 z_{\rm inc} 
\end{bmatrix},
\end{equation}
where $\mat{R}_y$ is the rotation matrix.
By performing the rotation transformation, one can easily obtain the expressions for TM- and TE-waves in the coordinate system $\mat{r}_{\rm inc} = (x_{\rm inc}, y_{\rm inc}, z_{\rm inc})$ in the following form~\cite{14}:
\begin{equation}\label{eq:3}
\begin{aligned}
&\mathbf{\Phi}_{\rm TM,TE}(\mat{k}_{\bot, {\rm inc}})\exp \left\{\ii \mat{k}_{\rm inc} \cdot \mat{r}_{\rm inc}\right\} 
\\&= 
\mathbf{\Phi}_{\rm TM,TE}(\mat{k}_{\bot, {\rm inc}})\exp \left\{
\ii k_{x, {\rm inc}} x_{\rm inc} + 
\ii k_{y, {\rm inc}} y_{\rm inc} + 
\ii k_{z, {\rm inc}} z_{\rm inc} \right\},
\end{aligned}
\end{equation}
where $\mat{k}_{\rm inc} = (k_{x,{\rm inc}}, k_{y,{\rm inc}}, k_{z,{\rm inc}}) = (\mat{k}_{\bot,{\rm inc}},k_{z,{\rm inc}})$ is the wave vector written in the coordinate system of the incident beam, $\mat{k}_{\bot, {\rm inc}} = (k_{x,{\rm inc}}, k_{y,{\rm inc}})$ is the vector of the transverse wave vector components, and the vectors $\mat{\Phi}_{\rm TM,TE}$ have the form
\begin{equation}\label{eq:4}
\mat{\Phi}_{\rm TM}(\mat{k}_{\bot, {\rm inc}}) = \frac{1}{{\left| \mat{k}_\bot \right|}}\left[{\begin{array}{*{20}{c}}
 \pmb{\alpha} \\ 
 \varepsilon_{\rm sup} \pmb{\beta} 
\end{array}} \right],\,\,{\mat{\Phi}}_{\rm TE}({\mat{k}_{\bot, {\rm inc}}}) = \frac{1}{\left|\mat{k}_\bot \right|}\left[{\begin{array}{*{20}{c}}
 -\pmb{\beta} \\ 
 \pmb{\alpha} 
\end{array}} \right],
\end{equation}
where
\begin{equation}\label{eq:5}
\left|\mat{k}_\bot \right| = \sqrt{k_{y,{\rm inc}}^2 + ({k_{x,{\rm inc}}}\cos\theta -{k_{z,{\rm inc}}}\sin\theta)^2},
\end{equation}
\begin{equation}\label{eq:6}
\begin{aligned}
 \pmb{\alpha} &= \begin{bmatrix}
 {-k_{x,{\rm inc}}k_{z,{{\rm inc}}}\cos\theta + (k_{y,{\rm inc}}^2 + k_{z,{\rm inc}}^2)\sin\theta} \\ 
 {-k_{y{{\rm, inc}}}({k_{z,{\rm inc}}}\cos\theta + k_{x,{\rm inc}}\sin\theta)} \\ 
 {(k_{y{{\rm, inc}}}^2 + k_{z,{\rm inc}}^2)\cos\theta -{k_{x,{\rm inc}}}{k_{z,{\rm inc}}}\sin\theta} 
\end{bmatrix}, 
\\
 \pmb{\beta} &= \begin{bmatrix}
 k_0 {k_{y,{\rm inc}}}\cos\theta \\ 
 k_0( -k_{x, {\rm inc}}\cos\theta + k_{z,{\rm inc}}\sin\theta) \\ 
 -k_0 k_{y, {\rm inc}} \sin\theta 
\end{bmatrix}. \\ 
\end{aligned}
\end{equation}
Note that the quantity $\left| \mat{k}_\bot \right|$ in Eq.~\eqref{eq:5} is the magnitude of the vector $\mat{k}_\bot = (k_x, k_y)$ obtained from the vector $\mat{k}_{\rm inc}$ by moving to the global coordinate system: $\mat{k}^{\rm T} = \mat{R}_y^{\rm T}\mat{k}_{\rm inc}^{{\rm T}}$.
Let us also note that the vectors $\mat{\Phi}_{\rm TM}$ and $\mat{\Phi}_{\rm TE}$ in Eqs.~\eqref{eq:3} and~\eqref{eq:4} are considered as functions of only the transverse components of the wave vector ${\mat{k}_{\bot, {\rm inc}}} = (k_{x,{\rm inc}},k_{y,{\rm inc}})$, since the remaining $z_{\rm inc}$ component can be expressed through them as ${k_{z,{\rm inc}}} = \pm \sqrt{k_0^2{\varepsilon_{\rm sup}} - k_{\bot, {\rm inc}}^2} $.
The signs ``$\pm$'' in the expression for $k_{z,{\rm inc}}$ correspond to the waves propagating in the positive and negative directions of the $z_{\rm inc}$ axis in the coordinate system associated with the incident beam, respectively.

Next, let us represent the incident optical beam propagating in the negative direction of the $z_{\rm inc}$ axis of the coordinate system ${{\mat{r}}_{\rm inc}} = ({x_{\rm inc}},{y_{\rm inc}},{z_{\rm inc}})$ as a superposition of TM- and TE-waves of Eqs.~\eqref{eq:3} and~\eqref{eq:4}:
\begin{equation}\label{eq:7}
\begin{aligned}
 \mat{P}_{\rm inc}(\mat{r}_{\rm inc}) = 
\iint & G_{\rm TE}(\mat{k}_{\bot, {\rm inc}})\mat{\Phi}_{\rm TE}({\mat{k}_{\bot, {\rm inc}}})
\\  &\cdot\exp \left\{\ii \mat{k}_{\bot, {\rm inc}} \mat{r}_{\bot, {\rm inc}} -\ii z_{\rm inc}\sqrt{k_0^2 \varepsilon_{\rm sup} - \mat{k}_{\bot, {\rm inc}}^2} \right\}\dd \mat{k}_{\bot, {\rm inc}} 
\\
 + \iint &G_{\rm TM}(\mat{k}_{\bot, {\rm inc}})\mat{\Phi}_{\rm TM}(\mat{k}_{\bot, {\rm inc}})
\\&\cdot\exp \left\{\ii\mat{k}_{\bot, {\rm inc}}\mat{r}_{\bot, {\rm inc}} - \ii{z_{\rm inc}}\sqrt{k_0^2 \varepsilon_{\rm sup} - \mat{k}_{\bot, {\rm inc}}^2} \right\}\dd \mat{k}_{\bot, {\rm inc}},
\end{aligned}
\end{equation}
where ${{\mat{r}}_{\bot, {\rm inc}}} = ({x_{\rm inc}},{y_{\rm inc}})$, ${G_{\rm TE}}({\mat{k}_{\bot, {\rm inc}}})$ and ${G_{\rm TM}}(\mat{k}_{\bot, {\rm inc}})$ are the spectra of TE- and TM-components of the incident beam, respectively, which represent the amplitudes of TE- and TM-polarized plane waves constituting the beam.
Note that since the incident beam propagates in the negative direction of the $z_{\rm inc}$ axis, the component $k_{z, {\rm inc}}$ in the wave vectors in the expressions for the vectors $\mat{\Phi}_{\rm TE}$, $\mat{\Phi}_{\rm TM}$ has to be taken with a minus sign, i.\,e., $k_{z, {\rm inc}} = -\sqrt{k_0^2 \varepsilon_{\rm sup} - \mat{k}_{\bot, {\rm inc}}^2}$.

In the general case, by choosing the spectra ${G_{\rm TE}}({\mat{k}_{\bot, {\rm inc}}})$ and ${G_{\rm TM}}(\mat{k}_{\bot, {\rm inc}})$, one can define two chosen components of the electromagnetic field.
In what follows, we will assume that in the plane $z_{\rm inc} = 0$, the incident beam is linearly polarized along the $x_{\rm inc}$ axis, so that $E_{{\rm inc},y}(x_{\rm inc}, y_{\rm inc}, 0) \equiv 0$, and the $x_{\rm inc}$ component of the electric field has a certain required form:
\begin{equation}\label{eq:8}
E_{{\rm inc},x}(x_{\rm inc}, y_{\rm inc}, 0) 
= \iint G_x({\mat{k}_{\bot, {\rm inc}}})\exp \{\ii\mat{k}_{\bot, {\rm inc}}\cdot\mat{r}_{\bot, {\rm inc}}\}\dd\mat{k}_{\bot, {\rm inc}},
\end{equation}
where $G_x(\mat{k}_{\bot, {\rm inc}})$ is the spectrum of this component, which is assumed to be given.
In this case, the spectra $G_{\rm TE}(\mat{k}_{\bot, {\rm inc}})$ and $G_{\rm TM}(\mat{k}_{\bot, {\rm inc}})$ can be easily obtained as~\cite{14}
\begin{equation}\label{eq:9}
\begin{aligned}
 G_{\rm TM}(\mat{k}_{\bot, {\rm inc}}) &= g_{\rm TM}(\mat{k}_{\bot, {\rm inc}}) G_x(\mat{k}_{\bot, {\rm inc}}),
 \\
 G_{\rm TE}(\mat{k}_{\bot, {\rm inc}}) &= g_{\rm TE}(\mat{k}_{\bot, {\rm inc}}) G_x(\mat{k}_{\bot, {\rm inc}}),
\end{aligned}
\end{equation}
where
\begin{equation}\label{eq:10}
\begin{aligned}
g_{\rm TM}(\mat{k}_{\bot, {\rm inc}}) &= \frac{-k_{x,{\rm inc}}\cos\theta + k_{z,{\rm inc}}\sin\theta}{{{k_{z,{\rm inc}}}\left|{{\mat{k}_\bot}} \right|}},
\\
g_{\rm TE}(\mat{k}_{\bot, {\rm inc}}) &= \frac{k_{y,{\rm inc}}({k_{x,{\rm inc}}}\sin\theta + {k_{z,{\rm inc}}}\cos\theta)}{k_0 k_{z,{\rm inc}}\left|{{\mat{k}_\bot}} \right|},
\end{aligned}
\end{equation}
and $\left|{{\mat{k}_\bot}} \right|$ is defined by Eq.~\eqref{eq:5}.
By substituting Eqs.~\eqref{eq:9} and~\eqref{eq:10} to Eq.~\eqref{eq:7}, we obtain the incident beam as
\begin{equation}\label{eq:11}
\begin{gathered}
 \mat{P}_{\rm inc}(\mat{r}_{\rm inc}) = \iint\left[{{g_{\rm TM}}(\mat{k}_{\bot, {\rm inc}})\mat{\Phi}_{\rm TM}(\mat{k}_{\bot, {\rm inc}}) +g_{\rm TE}(\mat{k}_{\bot, {\rm inc}})\mat{\Phi}_{\rm TE}(\mat{k}_{\bot,{\rm inc}})} \right] \\ 
 \cdot{G_x}({\mat{k}_{\bot, inc}},\omega )\exp \left\{{\ii{\mat{k}_{\bot, {\rm inc}}}{{\mat{r}}_{\bot,{\rm inc}}} -\ii{z_{\rm inc}}\sqrt{k_0^2\varepsilon_{\rm sup} -\mat{k}_{\bot, {\rm inc}}^2}} \right\}\dd\mat{k}_{\bot,{\rm inc}},
\end{gathered}
\end{equation}
where
\begin{equation}\label{eq:12}
g_{\rm TM}(\mat{k}_{\bot,{\rm inc}})\mat{\Phi}_{\rm TM}(\mat{k}_{\bot,{\rm inc}}) + g_{\rm TE}(\mat{k}_{\bot,{\rm inc}}){\mat{\Phi}_{\rm TE}}(\mat{k}_{\bot,{\rm inc}}) = 
\begin{bmatrix}
 1 \\ 
 0 \\ 
 k_{x,{\rm inc}}/k_{z,{\rm inc}} \\ 
 -k_{x,{\rm inc}} k_{y,{\rm inc}} / (k_0 k_{z,{\rm inc}}) \\ 
 (-k_{y,{\rm inc}}^2 + k_0^2 \varepsilon_{\rm sup}) / (k_0 k_{z,{\rm inc}}) \\ 
 -k_{y,{\rm inc}}/k_0 
\end{bmatrix}.
\end{equation}
Equations~\eqref{eq:11} and~\eqref{eq:12} show that in the incident beam, the $x_{\rm inc}$ component of the electric field $E_{{\rm inc},x}({x_{\rm inc}},{y_{\rm inc}},0)$ does indeed have the required form of Eq.~\eqref{eq:8}, and $E_{{\rm inc},y}({x_{\rm inc}},{y_{\rm inc}},0) \equiv 0$.

\subsection{Representation of the reflected beam}

When the incident beam is reflected from a layered structure, the amplitudes of TM- and TE-waves constituting the beam are multiplied by reflection coefficients ${R_{\rm TM}}(\mat{k}_\bot)$ and $R_{\rm TE}(\mat{k}_\bot)$, respectively, where $\mat{k}_\bot = (k_x, k_y)$ are the tangential components of the wave vector in the global coordinate system calculated from the equation $\mat{k}^{\rm T} = \mat{R}_y^{\rm T} \mat{k}_{\rm inc}^{\rm T}$.
From this equation, taking into account the negative sign of $k_{z,{\rm inc}}$, we obtain
\begin{equation}\label{eq:13}
\begin{aligned}
\mat{k}_\bot
&=
\mat{k}_\bot({\mat{k}_{\bot, {\rm inc}}}) 
\\&=
\left({k_{x,{\rm inc}}}\cos\theta + \sin\theta\sqrt{k_0^2 \varepsilon_{\rm sup} - k_{x,{\rm inc}}^2 - k_{y,{\rm inc}}^2}, \,\,\,{k_{y,{\rm inc}}}\right). \\ 
\end{aligned}
\end{equation}

In the coordinate system associated with the reflected beam ${{\mat{r}}_{\rm refl}} = ({{\mat{r}}_{\bot, {\rm refl}}},{z_{\rm refl}})$ (see Fig.~\ref{fig:1}), where $\mat{r}_{\bot, {\rm refl}} = (x_{\rm refl}, y_{\rm refl})$, the transverse components $\mat{k}_{\bot, {\rm inc}}$ of the wave vectors of the incident waves are not changed upon reflection, and the component $k_{z, {\rm inc}}$ changes sign, since the reflected beam propagates in the positive direction of the $z_{\rm refl}$ axis.
Therefore, taking into account Eq.~\eqref{eq:11}, the reflected beam (written in the coordinate system ${{\mat{r}}_{\rm refl}}$) can be represented as
\begin{equation}\label{eq:14}
\begin{aligned}
\mat{P}_{\rm refl}({{\mat{r}}_{\rm refl}}) = \iint &G_x({\mat{k}_{\bot, {\rm inc}}}){\mat{\Theta}}(\mat{k}_{\bot, {\rm inc}})
\\&\cdot\exp \left\{{\ii{\mat{k}_{\bot, {\rm inc}}}{{\mat{r}}_{\bot, {\rm refl}}} +\ii{z_{\rm refl}}\sqrt{k_0^2{\varepsilon_{\rm sup}} -\mat{k}_{\bot, {\rm inc}}^2}} \right\}\dd\mat{k}_{\bot, {\rm inc}},
\end{aligned}
\end{equation}
where
\begin{equation}\label{eq:15}
\begin{aligned}
\mat{\Theta}(\mat{k}_{\bot, {\rm inc}}) = &g_{\rm TM}({\mat{k}_{\bot, {\rm inc}}}){{\mat{\Phi}}_{\rm TM}}({\mat{k}_{\bot, {\rm inc}}}){R_{\rm TM}}({\mat{k}_\bot}(\mat{k}_{\bot, {\rm inc}})) 
\\&+g_{\rm TE}(\mat{k}_{\bot, {\rm inc}}){{\mat{\Phi}}_{\rm TE}}({\mat{k}_{\bot, {\rm inc}}}){R_{\rm TE}}({\mat{k}_\bot}({\mat{k}_{\bot, {\rm inc}}})).
\end{aligned}
\end{equation}
Here, the functions ${g_{\rm TM}}({\mat{k}_{\bot, {\rm inc}}})$ and ${g_{\rm TE}}({\mat{k}_{\bot, {\rm inc}}})$ are defined by Eq.~\eqref{eq:10}, and the vectors $\mat{\Phi}_{\rm TE}$, $\mat{\Phi}_{\rm TM}$ have the form of Eqs.~\eqref{eq:4}--\eqref{eq:6}, where ${k_{z,{\rm inc}}} = + \sqrt{k_0^2{\varepsilon_{\rm sup}} -\mat{k}_{\bot, {\rm inc}}^2} $.

From Eqs.~\eqref{eq:8},~\eqref{eq:14}, and~\eqref{eq:15}, it follows that the electromagnetic field components of the reflected beam at $z_{\rm refl} = 0$ correspond to the transformation of the $x_{\rm inc}$ component of the electric field of the incident beam ${E_{{{\rm inc,}}x}}({{\mat{r}}_{\bot, {\rm inc}}},0)$ by a linear system with a vectorial transfer function (TF) $\mat{\Theta}({\mat{k}_{\bot, {\rm inc}}})$ defined by Eq.~\eqref{eq:15}.

\section{Theoretical description of the vectorial differentiation}

In this section, we will obtain the conditions necessary to obtain spatial differentiation in different field components of the reflected beam.
Since conventional light detectors record the intensity (squared magnitude) of the electric field, further we will restrict our consideration to the electric field components of the incident and reflected beams.

Let us assume that the spectrum of the incident beam ${G_x}({\mat{k}_{\bot, {\rm inc}}})$ is “concentrated” in the vicinity of the point $\mat{k}_{\bot, {\rm inc}} = (0,0)$ [${\mat{k}_\bot}(0) = ({k_{x,0}},0)$, where $k_{x,0} = k_0\sqrt{\varepsilon_{\rm sup}} \sin\theta$], so that the TF of Eq.~\eqref{eq:15} in this vicinity can be approximated by its expansion into Taylor series up to linear terms.
By expanding in this way first three components of the TF of Eq.~\eqref{eq:15} describing the electric field of the reflected beam, we obtain:
\begin{equation}\label{eq:16}
\begin{bmatrix}
 \Theta_{E_x}(\mat{k}_{\bot,{\rm inc}}) \\ 
 \Theta_{E_y}(\mat{k}_{\bot,{\rm inc}}) \\ 
 \Theta_{E_z}(\mat{k}_{\bot,{\rm inc}}) 
\end{bmatrix}
 \approx 
\begin{bmatrix}
 {{R_{\rm TM}}({\mat{k}_\bot}(0))} \\ 
 0 \\ 
 0 
\end{bmatrix} 
+ \begin{bmatrix}
 c_x \\ 
 0 \\ 
 \frac{1}{k_0\sqrt{\varepsilon_{\rm sup}}}{R_{\rm TM}}(\mat{k}_\bot(0))
\end{bmatrix}{k_{x,{\rm inc}}} + 
\begin{bmatrix}
 0 \\ c_y \\ 0 
\end{bmatrix}{k_{y,{\rm inc}}},
\end{equation}
where
\begin{equation}\label{eq:17}
c_x = \cos \theta \frac{{\partial{R_{\rm TM}}}}{{\partial{k_{x,{\rm inc}}}}}({\mat{k}_\bot}(0))
,\;\;\; 
c_y = \frac{\cot\theta}{k_0\sqrt{{\varepsilon_{\rm sup}}}}\left[R_{\rm TM}({\mat{k}_\bot}(0)) -R_{\rm TE}(\mat{k}_\bot(0)) \right].
\end{equation}

Let us remind that the considered incident beam is linearly polarized along the $x_{\rm inc}$ axis so that $E_{{\rm inc},y}({x_{\rm inc}},{y_{\rm inc}},0) \equiv 0$.
However, as it can be seen from Eqs.~\eqref{eq:14}--\eqref{eq:16}, the cross-polarized component of the reflected beam ${E_{{\rm refl}, y}}({{\mat{r}}_{\bot, {\rm refl}}},0)$ becomes nonzero.
Moreover, it follows from Eq.~\eqref{eq:16} that the generation of the cross-polarized component ${E_{{\rm refl}, y}}({{\mat{r}}_{\bot, {\rm refl}}},0)$ can be described in the linear approximation by the following TF:
\begin{equation}\label{eq:18}
\Theta_{{\rm lin}, E_y}(\mat{k}_{\bot, {\rm inc}}) = c_y k_{y, {\rm inc}}.
\end{equation}
This TF, as one can easily obtain from Eq.~\eqref{eq:8}, describes the differentiation of the electric field component $E_{{\rm inc},x}$ of the incident beam with respect to the variable $y_{\rm inc}$:
\begin{equation}\label{eq:19}
E_{{\rm refl},y}({x_{\rm inc}},{y_{\rm inc}},0) = - \ii c_y \frac{\partial E_{{\rm inc},x}(x_{\rm inc}, y_{\rm inc},0)}{\partial y_{\rm inc}}.
\end{equation}
Note that in Refs.~\cite{24, 25}, the differentiation in the cross-polarized component, which occurs upon reflection of a linearly polarized beam from a layered structure (or just from a single interface) is considered as a manifestation of the optical analogue of the spin Hall effect.

In what follows, we will assume that the reflection coefficient of the layered structure $R_{\rm TM}(\mat{k}_\bot(0))$ vanishes.
In this case, in the linear approximation of Eqs.~\eqref{eq:16} and~\eqref{eq:17}, we obtain:
\begin{equation}\label{eq:20}
\Theta_{{\rm lin}, E_x}(\mat{k}_{\bot, {\rm inc}}) = 
c_x k_{x,{\rm inc}},
\,\,\,\,
\Theta_{{\rm lin}, E_z} (\mat{k}_{\bot, {\rm inc}}) = 0.
\end{equation}
From Eq.~\eqref{eq:8}, it is evident that the TF $\Theta_{{\rm lin}, E_x}(\mat{k}_{\bot, {\rm inc}})$ describes the differentiation of the electric field component of the incident beam $E_{{\rm inc},x}$  with respect to the variable $x_{\rm inc}$.
Thus, if the zero reflection condition $R_{\rm TM}(\mat{k}_\bot(0)) = 0$ is fulfilled, we will arrive at
\begin{equation}\label{eq:21}
\begin{gathered}
 E_{{\rm refl},x}(x_{\rm inc}, y_{\rm inc}, 0) = - \ii{c_x}\frac{\partial E_{{\rm inc},x}(x_{\rm inc}, y_{\rm inc}, 0)}{\partial x_{\rm inc}}, \hfill \\
 E_{{\rm refl},z}({x_{\rm inc}},{y_{\rm inc}},0) = 0. \hfill \\ 
	\end{gathered}
\end{equation}
As it was mentioned above, the reflection zeros are often associated with the excitation of eigenmodes of the layered structure~\cite{11,12}.
In this regard, the obtained Eqs.~\eqref{eq:19} and~\eqref{eq:21} demonstrate that by simultaneously utilizing the optical analogue of the spin Hall effect and a resonant effect providing a reflection zero, one can perform the operation, which can be referred to as the vectorial differentiation and consists in computing the partial derivatives of the $x_{\rm inc}$ component of the linearly polarized incident beam with respect to two different variables (${x_{\rm inc}}$ and ${y_{\rm inc}}$) in two different electric field components of the reflected beam ($E_{{\rm refl},x}$ and $E_{{\rm refl},y}$).
In our opinion, the most interesting case from the practical point of view is the so-called “isotropic differentiation”, which is performed at $\left|{{c_x}} \right| = \left|{{c_y}} \right|$. 
If $R_{\rm TM}(\mat{k}_\bot(0)) = 0$, this condition reads as
\begin{equation}\label{eq:22}
\cos\theta \cdot \left|\frac{\partial R_{\rm TM}}{\partial k_{x,{\rm inc}}}({\mat{k}_\bot}(0)) \right| 
= \frac{\cot\theta}{k_0\sqrt{\varepsilon_{\rm sup}}} \cdot \left|{R_{\rm TE}}({\mat{k}_\bot}(0)) \right|.
\end{equation}
In this case, the magnitude of the vectorial TF of Eq.~\eqref{eq:16} will depend on $\left| \mat{k}_{\bot,{\rm inc}} \right| = \sqrt{k_{x,{\rm inc}}^2 + k_{y,i{{\rm nc}}}^2} $, and the intensity of the reflected beam (i.\,e., the squared magnitude of the electric field $\left| \mat{E}_{\rm refl} \right|^2 =\left|E_{{\rm refl},x} \right|^2 + \left| E_{{\rm refl},y} \right|^2 + \left| E_{{\rm refl}, z} \right|^2$) will be proportional to the squared absolute value of the gradient of the electric field of the incident beam:
\begin{equation}\label{eq:23}
\begin{aligned}
I_{\rm refl}(x_{\rm inc}, y_{\rm inc}) 
&= |c_x|^2 \left({{\left|{\frac{\partial E_{{\rm inc},x}(x_{\rm inc}, y_{\rm inc}, 0)}{\partial x_{\rm inc}}} \right|}^2} + \left|{\frac{\partial E_{{\rm inc},x}(x_{\rm inc}, y_{\rm inc}, 0)}{\partial y_{\rm inc}}} \right|^2 \right) 
\\&= |c_x|^2 \left|\nabla E_{{\rm inc},x}(x_{\rm inc}, y_{\rm inc}, 0) \right|^2.
\end{aligned}
\end{equation}
In the opinion of the present authors, the implementation of the transformation defined by Eq.~\eqref{eq:23} is of great practical interest, since it enables isotropic edge detection (contour enhancement) in input images.

To complete this section, let us note that the obtained formulas~\eqref{eq:16}--\eqref{eq:23} are not applicable in the case of normal incidence (i.\,e., at $\theta = 0$).
This is due to the fact that in the vicinity of normal incidence, the reflection coefficients are even functions with respect to $k_x, k_y$ and therefore have a quadratic form:
\begin{equation}\label{eq:24}
R_{\rm TE,TM}(k_x, k_y) 
=
R_{\rm TE,TM}\left(\sqrt{k_x^2 + k_y^2}, 0\right)
\approx r_0 + r_{\rm 2,TE,TM} (k_x^2 + k_y^2).
\end{equation}
where $r_0 = R_{\rm TE}(0,0) = R_{\rm TM}(0,0)$, $r_{\rm 2,TE,TM} = \frac{\partial^2{R_{\rm TE,TM}}}{\partial k_x^2}(0,0)$.

Assuming that $\theta = 0$, we have $\mat{k}_\bot = \mat{k}_{\bot, {\rm inc}}$ in the expressions defining the vectorial TF [Eq.~\eqref{eq:15}].
In this case, using the expansion of Eq.~\eqref{eq:24} at $r_0 = 0$, it is easy to obtain that at normal incidence, the transformation of the $x$ component of the electric field (in the quadratic approximation) is described by the TF
\begin{equation}\label{eq:25}
\Theta_{{\rm quad}, E_x}(\mat{k}_\bot) = {r_{2,TM}}\,k_x^2 +{r_{2,TE}}\,k_y^2,
\end{equation}
which corresponds to the following second-order differentiation:
\begin{equation}\label{eq:26}
E_{{\rm refl}, x}(x,y,0) 
= -{r_{2,TM}}\frac{{{\partial^2}{E_{{\rm inc},x}}(x,y,0)}}{{\partial{x^2}}} -{r_{2,TE}}\frac{{{\partial^2}E_{{\rm inc},x}(x,y,0)}}{{\partial{y^2}}}.
\end{equation}

It is interesting to note that the formation of the cross-polarized component of the electric field of the reflected beam in this case is described by the following TF:
\begin{equation}\label{eq:27}
\Theta_{{\rm quad}, E_y}(\mat{k}_\bot) = k_x k_y\left( r_{\rm 2,TM} - r_{\rm 2,TE} \right),
\end{equation}
which corresponds to the computation of the second-order mixed partial derivative:
\begin{equation}\label{eq:28}
E_{{\rm refl}, y}(x,y,0) = \left( r_{\rm 2,TE} - r_{\rm 2,TM} \right)\frac{\partial^2 E_{{\rm inc},x}(x,y,0)}{\partial x\partial y}.
\end{equation}
In recent paper~\cite{26}, the transformation of Eq.~\eqref{eq:28} was considered as a manifestation of the high-order spin Hall effect of light.
Thus, in the normal incidence geometry, it is possible to compute a weighted sum of the second derivatives in the component $E_{{\rm refl}, x}$ and the second mixed derivative in the cross-polarized component $E_{{\rm refl}, y}$.
Despite the fact that the second-order differentiation also enables enhancing contours on images, in what follows, we will limit our consideration to the case of oblique incidence and the corresponding isotropic vectorial differentiation operation of Eq.~\eqref{eq:23}, which, in our opinion, is more interesting from the practical point of view.

\section{Three-layer metal-dielectric structure for vectorial differentiation}
According to the conditions of vectorial differentiation obtained in the previous section for the oblique incidence geometry, the layered structure must possess a reflection zero for an incident TM-polarized plane wave with the tangential wave vector components $\mat{k}_\bot(0) = (k_{x,0},0) = (k_0\sqrt{{\varepsilon_{\rm sup}}} \sin \theta, 0)$.
In addition, to obtain isotropic differentiation described by Eq.~\eqref{eq:23}, it is necessary for the reflection coefficients of the structure to provide the fulfillment of the equality $|c_x| = |c_y|$ defined by Eq.~\eqref{eq:22} and relating the derivative $\frac{\partial R_{\rm TM}}{\partial k_{x,{\rm inc}}}(\mat{k}_\bot(0))$ and the reflection coefficient for the cross-polarized wave $R_{\rm TE}(\mat{k}_\bot(0))$.
Therefore, the utilized layered structure not only has to possess a reflection zero ${R_{\rm TM}}({\mat{k}_\bot}(0)) = 0$, but must also have a certain additional free parameter for being able to fulfill the isotropic differentiation condition of Eq.~\eqref{eq:22}.
The authors believe that the simplest layered structure, which can satisfy the discussed requirements, is the three-layer metal-dielectric-metal (MDM) structure schematically shown in Fig.~\ref{fig:1} and consisting of an upper metal layer (with thickness $h_{\rm m,1}$ and dielectric permittivity $\varepsilon_{\rm m,1}$), a dielectric layer (thickness $h_d$, dielectric permittivity $\varepsilon_d$), and a lower metal layer (thickness $h_{\rm m,2}$, dielectric permittivity $\varepsilon_{\rm m,2}$).
The considered MDM structure is resonant since it supports leaky modes localized in the dielectric layer~\cite{11,12}.
At the so called critical coupling conditions, the reflection coefficient of the structure can vanish.
Moreover, in~\cite{11,12}, it was shown that at fixed materials (values of $\varepsilon_{\rm m, 1}$, $\varepsilon_{\rm m, 2}$, and $\varepsilon_{\rm d}$) and the parameters of the incident wave (wavelength $\lambda $, angle of incidence $\theta $, and fixed TM- or TE-polarization), it is always possible to obtain a reflection zero in the MDM structure by choosing the thicknesses $h_{\rm m,1}$ and $h_{\rm d}$ of the upper metal and the dielectric layer.
It is important to note that the thickness $h_{\rm m,2}$ of the lower metal layer of the MDM structure can be considered as a free parameter, i.\,e., for different thicknesses of the lower layer, it is possible to obtain different MDM structures with a reflection zero ${R_{\rm TM}}(\mat{k}_\bot(0)) = 0$ and different values of $|c_x|$, $|c_y|$.

As an example, we considered Cu--${\rm TiO}_2$--Cu MDM structures (i.\,e., structures consisting of two copper layers separated by a layer of titanium dioxide) on a ${\rm TiO}_2$ substrate (see Fig.~\ref{fig:1}).
The possibility to fulfill the condition of Eq.~\eqref{eq:22} was checked for MDM structures with different thicknesses of the lower metal layer $h_{\rm m,2}$ possessing reflection zeros $R_{\rm TM}(\mat{k}_\bot(0)) = 0$ at the wavelength $\lambda=633\nm$ and angle of incidence $\theta = 45^\circ$.
For the dielectric permittivities of the materials of the structure, reference data were used~\cite{31}.
The thicknesses of the upper two layers of the MDM structures providing reflection zero (at the chosen $h_{\rm m,2}$) were calculated using the previously developed numerical approach~\cite{12}.
As a result, an MDM structure was found, for which $c_x = 0.3472\exp\{2.8729\ii\}$, $c_y = 0.3467\exp\{0.8779\ii\}$ and, consequently, the required equality $|c_x| = |c_y|$ is fulfilled with a high accuracy.
The layer thicknesses of the found structure amount to
\begin{equation}\label{eq:29}
h_{\rm m,1} = 30.6\nm,\,\,h_{\rm d} = 68.9\nm,\,\,h_{\rm m,2} = 65.5\nm.
\end{equation}

\begin{figure}[hbt]
	\centering
		\includegraphics{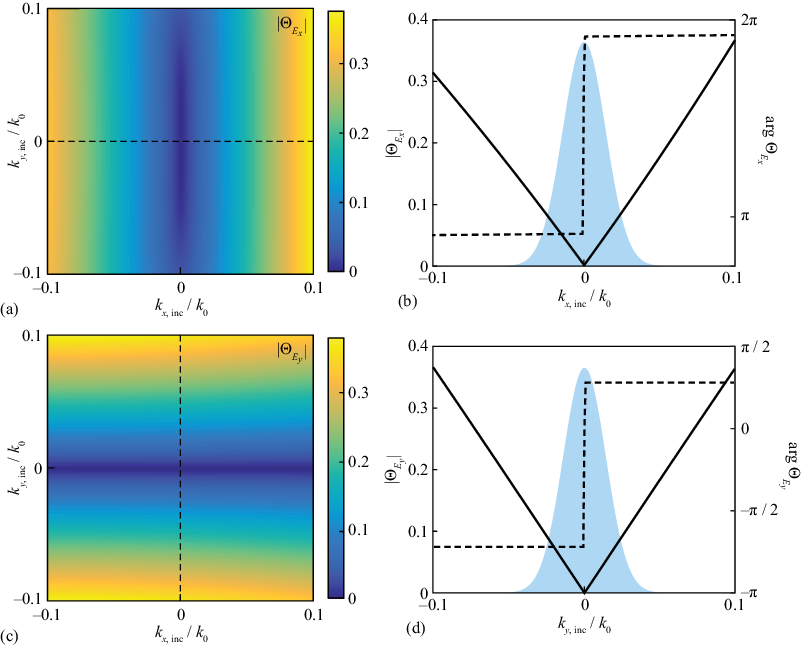}
	\caption{\label{fig:2} (a), (c) Absolute values of the calculated transfer functions $\Theta_{E_x}(\mat{k}_{\bot, {\rm inc}})$ (a) and $\Theta_{E_y}(\mat{k}_{\bot, {\rm inc}})$ (c) of the MDM structure of Eq.~\eqref{eq:29}; (b), (d) central cross-sections of the absolute values (solid lines, left vertical axes) and arguments (dashed lines, right vertical axes) of the TFs. Filled regions in (b) and (d) show the cross-sections of the spectrum of the incident Gaussian beam.}
\end{figure}

Figure~\ref{fig:2} shows the absolute values of the TFs $\Theta_{E_x}({\mat{k}_{\bot, {\rm inc}}})$ and $\Theta_{E_y}(\mat{k}_{\bot, {\rm inc}})$ for the structure of Eq.~\eqref{eq:29} [Figs.~\ref{fig:2}(a) and~\ref{fig:2}(c)] calculated using the “rigorous” Eq.~\eqref{eq:15} and the central cross-sections of their absolute values and arguments [Figs.~\ref{fig:2}(b) and~\ref{fig:2}(d)].
Let us note that the reflection coefficients, through which the TFs are expressed, were calculated using the numerically stable enhanced transmittance matrix approach~\cite{32}.
It is evident that in the considered spatial frequency range $|k_{x,{\rm inc}}| \leqslant 0.1k_0$, $|k_{y,{\rm inc}}| \leqslant 0.1k_0$, the absolute values of the calculated TFs are very close to their linear approximations $\Theta_{{\rm lin}, E_x}(\mat{k}_{\bot, {\rm inc}}) = c_x k_{x, {\rm inc}}$ and $\Theta_{{\rm lin}, E_y}(\mat{k}_{\bot, {\rm inc}}) = c_y k_{y, {\rm inc}}$ (the normalized root-mean-square deviation of the TFs from these approximations does not exceed 0.5\%).
Let us also note that using the rigorous Eq.~\eqref{eq:15}, the TF $\Theta_{E_z}(\mat{k}_{\bot, {\rm inc}})$ was also calculated, which describes the formation of the $z_{\rm refl}$ component of the reflected beam and is equal to zero in the linear approximation [see Eq.~\eqref{eq:20}].
It was obtained that in the spatial frequency range of interest, the magnitude of this TF is negligibly small in comparison with the TFs $\Theta_{E_x}(\mat{k}_{\bot, {\rm inc}})$ and $\Theta_{E_y}(\mat{k}_{\bot, {\rm inc}})$, therefore, it is not presented in the paper for the sake of brevity.
At the same time, this TF was taken into account in all calculations presented in the next section.

\section{Numerical investigation of the designed MDM structure}

Next, let us investigate the implementation of the vectorial differentiation operation by considering an incident linearly polarized three-dimensional Gaussian beam with the following $x_{\rm inc}$ component of the electric field:
\begin{equation}\label{eq:30}
E_{{\rm inc},x}(x_{\rm inc}, y_{\rm inc}, 0) = \exp \left\{- \frac{x_{\rm inc}^2 + y_{\rm inc}^2}{\sigma^2} \right\}.
\end{equation}
The spatial spectrum of this component is also described by a Gaussian function:
\begin{equation}\label{eq:31}
G_x(\mat{k}_{\bot, {\rm inc}}) =
 \frac{\sigma^2}{4\pi} \exp \left\{-\frac{\sigma^2}{4} (k_{x, {\rm inc}}^2 + k_{y, {\rm inc}}^2)\right\}.
\end{equation}

In Fig.~\ref{fig:2}, filled regions show the cross-sections of the spectrum $G_x(\mat{k}_{\bot, {\rm inc}})$ at $\sigma = 10 \um$.
Since the spectrum of the beam lies in the linearity interval of the TFs $\Theta_{E_x}(\mat{k}_{\bot, {\rm inc}})$ and $\Theta_{E_y}(\mat{k}_{\bot, {\rm inc}})$, one should expect high-quality computation of the derivatives in the considered example.

\begin{figure}[hbt]
	\hspace{-5em}
		\includegraphics{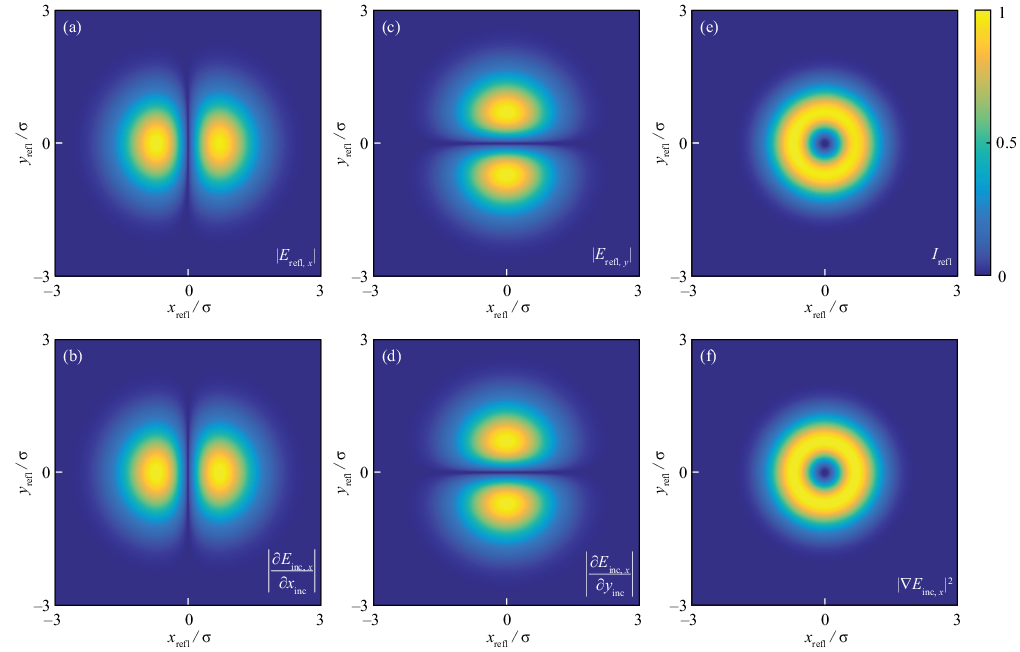}
	\caption{\label{fig:3} (a), (c), (e) Absolute values of the electric field components (a), (c) and intensity (e) of the reflected beam. (b), (d), (f) Absolute values of the analytically calculated derivatives $\left|\partial E_{{\rm inc},x} / \partial x_{\rm inc} \right|$ (b), $\left|{\partial E_{{\rm inc},x}} / \partial y_{\rm inc} \right|$ (d) and squared gradient $\left|\nabla E_{{\rm inc},x} \right|^2$ (f). All presented distributions are normalized by maximum values.
}
\end{figure}

In the upper row of Fig.~\ref{fig:3}, the absolute values of the $x_{\rm refl}$ and $y_{\rm refl}$ components of the electric field of the reflected beam numerically calculated using Eqs.~\eqref{eq:14} and~\eqref{eq:15} are shown, as well as the intensity of the reflected beam $I_{{\rm refl},y} = |E_{{\rm refl},x} |^2 + |E_{{\rm refl}, y} |^2 + |E_{{\rm refl}, z} |^2$ in the plane $z_{\rm refl} = 0$.
In the lower row of this figure, the absolute values of the analytically calculated partial derivatives 
$\left|\partial E_{{\rm inc},x} / \partial x_{\rm inc} \right|$, 
$\left|\partial E_{{\rm inc},x} / \partial y_{\rm inc} \right|$ and the squared absolute value of the gradient 
$\left|\nabla E_{{\rm inc},x} \right|^2 = 
\left|\partial E_{{\rm inc},x} / \partial x_{\rm inc} \right|^2 +
\left|\partial E_{{\rm inc},x} / \partial y_{\rm inc} \right|^2$ are shown. 
For ease of comparison, all the distributions presented in Fig.~\ref{fig:3} are normalized by maximum values. 
It is evident that the numerical simulation results [Figs.~\ref{fig:3}(a),~\ref{fig:3}(c), and~\ref{fig:3}(e)] are in very good agreement with the analytically calculated derivatives [Figs.~\ref{fig:3}(b),~\ref{fig:3}(d), and~\ref{fig:3}(f)] and, consequently, with the developed theoretical description [Eqs.~\eqref{eq:21},~\eqref{eq:19}, and~\eqref{eq:23}]. 
In particular, the normalized root-mean-square deviation of the numerically calculated intensity of the reflected field [Fig.~\ref{fig:3}(e)] from the analytically calculated squared gradient of the $x_{\rm inc}$ component of the electric field of the incident beam [Fig.~\ref{fig:3}(f)] amounts to only 0.39\%.

\begin{figure}[hbt]
	\centering
		\includegraphics{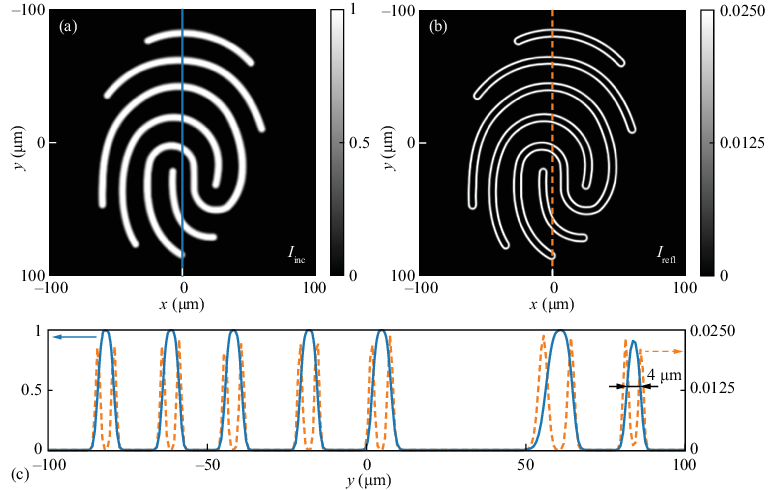}
	\caption{\label{fig:4} Intensity of the incident beam corresponding to an image of a fingerprint~(a) and numerically calculated intensity of the reflected beam~(b). (c)~Intensity cross-sections of the incident (solid line) and reflected (dashed line) beams along the vertical lines in~(a) and~(b).
}
\end{figure}

Let us now investigate the possibility of using the designed MDM structure for optical edge detection. 
As an example, we consider a linearly polarized incident beam with the intensity (squared absolute value of the $x_{\rm inc}$ component of the electric field) shown in Fig.~\ref{fig:4}(a) and corresponding to an image of a fingerprint. 
The normalized profile of the intensity distribution at $x_{\rm inc} = 0$ is presented with a solid line in Fig.~\ref{fig:4}(c), which shows that the thickness of the papillary lines on the input fingerprint image is of about $4\um$. 
Figure~\ref{fig:4}(b) shows the intensity of the reflected beam numerically calculated using the “rigorous” Eqs.~\eqref{eq:14} and~\eqref{eq:15}.
The spectrum $G_x(\mat{k}_{\bot,{\rm inc}})$ of the $x_{\rm inc}$ component of the incident beam was also calculated using numerical integration. Figure~\ref{fig:4}(b) demonstrates high-quality detection of edges with different orientation. 
The maximum intensity of the reflected beam is of about 2.5\% of the maximum intensity of the incident beam. 
In addition to the intensity profile of the incident beam, the dashed line in Fig.~\ref{fig:4}(c) shows the corresponding profile of the reflected beam. 
In Fig.~\ref{fig:4}(c), the formation of sharp intensity maxima at the edges of the lines of the input distribution is clearly visible.
Let us note that the proposed structure generates a single contour line for each edge; 
this behaviour is different from the edge-detecting structures based on computing the Laplace operator~\cite{3,14,29}, where ``doubled'' contour lines are produced for each edge.

\section{Conclusion}
In the present work, we considered the diffraction of a three-dimensional optical beam on a layered structure. 
We obtained a vectorial transfer function describing the transformation of electromagnetic field components of a linearly polarized incident beam occurring upon its reflection from the structure. 
Using this transfer function, we theoretically described the vectorial differentiation operation consisting in the computation of two partial derivatives with respect to two spatial coordinates in different transverse electric field components of the reflected beam. 
We demonstrated that the implementation of the vectorial differentiation is based on simultaneous utilization of the effect of optical resonance providing a reflection zero and of the optical analogue of the spin Hall effect. 
As an example of a layered structure implementing this operation, we proposed a three-layer MDM structure operating in the oblique incidence geometry. 
It was shown that by choosing the parameters of the MDM structure, one can achieve the so-called isotropic vectorial differentiation. 
In this case, the intensity of the reflected light beam (squared magnitude of the electric field) is proportional to the squared absolute value of the gradient of the electric field of the linearly polarized incident beam. 
The presented numerical simulation results of the designed MDM structure demonstrate high-quality vectorial differentiation as well as good performance of the structure in the problem of optical edge detection.

The obtained results may find application in the creation of systems of analog optical computing and optical information processing.

\section{Acknowledgments}
This work was funded by the Russian Science Foundation (project no. 24-12-00028, theoretical description of the vectorial differentiation, design and numerical investigation of the “isotropic” differentiator based on an MDM structure) and performed within the state assignment of NRC ``Kurchatov Institute'' (development of software for simulating diffraction of an optical beam on a layered structure).

\bibliographystyle{elsarticle-num} 
\bibliography{VDiff}

\end{document}